\documentclass[10pt,prd,aps,twocolumn,nofootinbib,superscriptaddress,showpacs,floatfix,amssymb,amsmath]{revtex4-1}
\usepackage{slashed}
\usepackage{epsfig}
\usepackage{multirow}

\newcommand{\beqn}{\begin{eqnarray}}
\newcommand{\eeqn}{\end{eqnarray}}
\newcommand{\eq}[1]{(\ref{#1})}

\newcommand{\cL}{{\cal L}}

\newcommand{\cZ}{{\cal Z}}

\newcommand{\cD}{{\cal D}}

\newcommand{\cl}{{\mathrm{cl}}}
\newcommand{\Tr}{{\mathrm{Tr}\,}}

\newcommand{\measure}{{\mathrm{measure}}}
\newcommand{\action}{{\mathrm{action}}}

\newcommand{\QED}{{\mathrm{QED}}}

\newcommand{\bs}{\boldsymbol}
\newcommand{\bra}[1]{{\left\langle#1\right|}}
\newcommand{\ket}[1]{{\left|#1\right\rangle}}
\newcommand{\avr}[1]{{\left\langle #1 \right\rangle}}
\def\bbbone{{\mathchoice {\rm 1\mskip-4mu l} {\rm 1\mskip-4mu l} {\rm 1\mskip-4.5mu l} {\rm 1\mskip-5mu l}}}
\newcommand{\dxp}{\overset{\leftrightarrow}{\partial}}

\usepackage{color}
\definecolor{purple}{rgb}{0.8,0,0.6}
\newcommand{\revision}[1]{{#1}}

\newcommand{\revisionZZ}[1]{{#1}}

\begin{document}

\title{Scale Magnetic Effect in Quantum Electrodynamics and the Wigner-Weyl Formalism}

\author{M.~N.~Chernodub }
\affiliation{Laboratoire de Math\'ematiques et Physique Th\'eorique UMR 7350, Universit\'e de Tours, 37200 France}
\affiliation{Laboratory of Physics of Living Matter, Far Eastern Federal University, Sukhanova 8, Vladivostok, 690950 Russia}
\author{M.~A.~Zubkov}
\affiliation{ITEP, B. Cheremushkinskaya 25, Moscow, 117259 Russia}
\affiliation{Moscow Institute of Physics and Technology, 9, Institutskii per., Dolgoprudny, Moscow Region, 141700, Russia}

\date{July 26, 2017}

\begin{abstract}
The Scale Magnetic Effect (SME) is the generation of electric current due to conformal anomaly in external magnetic field in curved spacetime. The effect appears in a vacuum with electrically charged massless particles. Similarly to the Hall effect, the direction of the induced anomalous current is perpendicular to the direction of the external magnetic field $\bs B$ and to the gradient of the conformal factor $\tau$, while the strength of the current is proportional to the beta function of the theory. In massive electrodynamics the SME remains valid, but the value of the induced current differs from the current generated in the system of massless fermions. In the present paper we use the Wigner--Weyl formalism to demonstrate that in accordance with the decoupling property of heavy fermions the corresponding anomalous conductivity vanishes in the large-mass limit with $m^2 \gg |e {\bs B}|$ and $m \gg |\nabla \tau|$.
\end{abstract}

\maketitle

\section{Introduction}

Anomalous transport phenomena have attracted wide attention of the scientific community in recent years~\cite{Basar:2012gm,Landsteiner:2012kd}. The anomalous transport is associated with quantum anomalies~\cite{Shifman:1988zk} which break original symmetries of classical systems due to quantum fluctuations. The axial anomaly and the mixed axial-gravitational anomaly are suggested to lead to various transport phenomena such as the Chiral Magnetic~\cite{Fukushima:2008xe,Vilenkin:1980fu}, Chiral Separation~\cite{Son:2004tq,Metlitski:2005pr} and Chiral Vortical~\cite{Vilenkin:1979ui,Son:2009tf} Effects that generate both electric (vector) and axial (pseudovector) currents as well as energy flows~\cite{Landsteiner:2011cp,Braguta:2013loa} in usual or in chirally imbalanced matter. These currents and flows may be directed along the axis of a background magnetic field or along a vorticity vector in the case if the matter is rotating. The anomalous transport appears both in solid state~\cite{Volovik2003} and particle~\cite{Kharzeev:2015kna} physics contexts.

The basic rule is that anomalous symmetry breaking may be associated with a certain (anomalous) transport law that cannot otherwise appear in a classical system with an unbroken classical symmetry. Besides the axial anomalies, certain theories may also exhibit the conformal anomaly associated with breaking of the classical conformal invariance at the quantum level. In Ref.~\cite{Chernodub:2016lbo} it was indeed shown that the anomalous breaking of conformal (scale) symmetry in conformally invariant gauge theories should also lead to emergence of two new transport phenomena, the scale magnetic effect (SME) and the scale electric effect (SEE), which generate electric current in electromagnetic field background in curved spacetime. The SME is a stationary phenomenon which induces an electric current perpendicularly to the direction of external magnetic field in a static curved space. The SEE is a nonstationary effect which is realized in an external electric field in time-dependent gravitational background. The generated electric currents are proportional to a beta function of the corresponding theory. The explicit expressions for the SME and SEE are given, respectively, in Eqs.~\eq{eq:SME} and \eq{eq:SEE} below. Both these effects appear in the theory with vanishing fermion masses or with nearly vanishing fermion masses that are much smaller than the energy scales associated with both external electric/magnetic field and the variations of the gravitational field.

The aim of the present article is to consider the opposite limit when the fermion mass is much larger than the energy scales given by the external electromagnetic field and the gradient of the gravitational field. Firstly, we notice that the SME and SEE phenomena are associated with the contribution of the conformal anomaly to the trace of the energy-momentum tensor in a classical electromagnetic field background~\cite{Chernodub:2016lbo}. In the (classically conformal) massless case this trace is entirely given by an anomalous contribution which originates from the change of the integration measure with respect to the Weyl transformation~\cite{Fujikawa:1980vr}. In the massive case the trace of the energy--momentum tensor contains also a non-anomalous extra contribution emerging due to the explicit breaking of the conformal symmetry in the classical Lagrangian. In our article we calculate the non-anomalous contribution to the scale magnetic effect in QED using the Wigner-Weyl formalism. We demonstrate that in the limit of large fermion mass the non-anomalous contribution cancels precisely the anomalous contribution coming from the integration measure over the fermionic fields. Therefore, in agreement with decoupling theorems, the scale magnetic effect is strongly suppressed for sufficiently massive fermions.

The structure of the paper is as follows. The next three sections are devoted to brief reviews of the scale magnetic and electric effects (Sect.~\ref{eq:SME:review}), relevant features of QED in a curved spacetime (Sect.~\ref{eq:curved}) and basics of the Wigner-Weyl approach (Sect.~\ref{ref:Wigner}). In Section~\ref{ref:current} we derive the non-anomalous part of the electric current of SME in QED with massive fermions, and demonstrate that it cancels precisely the contribution of integration measure. The last section is devoted to discussion of our results.

\section{Scale Magnetic/Electric Effects}

\label{eq:SME:review}

\subsection{Masless fermions}

Let us consider massless QED with one species of a Dirac fermion $\psi$ in (3+1) spacetime dimensions:
\beqn
\cL = - \frac{1}{4} F^{\mu\nu} F_{\mu\nu}  + {\bar \psi} i \gamma^\mu D_\mu \psi\,,
\label{eq:L:QED}
\eeqn
where $\gamma^\mu$ are the Dirac matrices, $F_{\mu\nu} = \partial_\mu A_\nu -  \partial_\nu A_\mu$ is the field strength tensor of the gauge field $A_\mu$ and $D_\mu = \partial_\mu + i e A_\mu$ is the covariant derivative.

In the massless QED the trace of the classical stress-energy tensor $T^{\mu\nu}$ is identically zero, $(T^\mu_\mu)_{\cl} \equiv 0$, because of the conformal invariance of the theory at the classical level. The theory does not contain a characteristic energy scale thus implying the invariance  of the classical action $S = \int d^4 x \, \cL$ under the scale transformations: $x \to \lambda^{-1} x$, $A_\mu \to \lambda A_\mu$ and $\psi \to \lambda^{3/2} \psi$. However quantum corrections make the electric charge $e$ dependent on the renormalization energy scale, $e = e(\mu)$. The apparent non-invariance of the quantum theory on the energy scale is explicitly manifested in a nonzero beta-function of the theory:
\beqn
\beta(e) = \frac{{\mathrm d} e}{{\mathrm d} \ln \mu}\,.
\label{eq:beta}
\eeqn
As a result, in the background of the classical electromagnetic field $A^{\cl}_\mu$ the trace of the stress-energy tensor becomes nonzero due to quantum corrections~\cite{Shifman:1988zk}:
\beqn
\avr{T^\alpha_\alpha(x)} = \frac{\beta(e)}{2 e}  F^{\cl,\mu\nu}(x) F^\cl_{\mu\nu}(x).
\label{eq:Tmunu:avr}
\eeqn
Below we study the effects in classical background gauge fields only and therefore we omit hereafter the superscript ``cl''  in~$A^{\cl}_\mu$ and $F^\cl_{\mu\nu}$.

A simplest way to reveal anomalous transport effects emerging due to the conformal (scale) anomaly~\eq{eq:Tmunu:avr} is to consider the following conformally flat metric:
\beqn
g_{\mu\nu}(x) = e^{2 \tau(x)} \eta_{\mu\nu}\,,
\label{eq:g:munu}
\eeqn
where $\eta_{\mu\nu} = {\mathrm{diag}}(+1,-1,-1,-1)$ is the flat Minkowski metric. Starting from Eq.~\eq{eq:Tmunu:avr}, and assuming that the conformal factor in Eq.~\eq{eq:g:munu} is small $|\tau| \ll 1$, one can show that in the curved background \eq{eq:g:munu} the conformal (scale) anomaly generates an anomalous electric current in the presence of the background of magnetic field $\bs B$:
\beqn
{\bs J} & = &  \frac{2 \beta(e)}{e} {\bs \nabla} \tau(x) \times {\bs B}(x)\,,
\label{eq:SME}
\eeqn
which is proportional to the gradient of the local scale factor $\tau(x)$ of the conformally flat metric~\eq{eq:g:munu}. The anomalous generation of the electric current by background magnetic field~\eq{eq:SME} is the scale magnetic effect (SME) proposed in Ref.~\cite{Chernodub:2016lbo}.

In the electric field background $\bs E$ the anomalous generation of the electric current resembles the Ohm law~\cite{Chernodub:2016lbo}:
\beqn
{\bs J} = \sigma(x) {\bs E}(x)\,,
\label{eq:SEE}
\eeqn
with the essential difference that the metric-dependent anomalous electric conductivity
\beqn
\sigma(t, {\bs x}) & = & - \frac{2 \beta(e)}{e} \frac{\partial \tau(t,{\bs x})}{\partial t}\,,
\label{eq:sigma:anomalous}
\eeqn
may take {\emph{negative}} values. Equations~\eq{eq:SEE} and \eq{eq:sigma:anomalous} determine the scale electric effect (SEE). It was suggested that the SEE describes the negative vacuum conductivity associated with the Schwinger pair production in an expanding de Sitter Universe. Earlier, a negative electric conductivity has indeed been found for fermionic~\cite{Hayashinaka:2016qqn} and bosonic~\cite{Kobayashi:2014zza} Schwinger effects.

Notice that the classical electric current induced by the external electromagnetic field in the conformal background~\eq{eq:g:munu} is identically zero. The scale magnetic~\eq{eq:SME} and scale electric~\eq{eq:SEE} effects are related to each other as they are originating from the same Lorentz-covariant expression~\cite{Chernodub:2016lbo}. The corresponding currents are proportional to the beta function~\eq{eq:beta}. Below we rederive the SME current~\eq{eq:SME} using a straightforward calculation based on a truncated Wigner expansion. This approach will also allows us to identify possible effects of a nonzero fermion mass on the anomalous current.

\subsection{Massive fermions}

QED with one species of massive Dirac fermions is described by the following Lagrangian:
\beqn
\cL = - \frac{1}{4} F^{\mu\nu} F_{\mu\nu}  + {\bar \psi} \Big(i \gamma^\mu D_\mu - m \Big) \psi\,.
\label{eq:L:QEDm}
\eeqn
With the help of the perturbative methods the trace of the corresponding energy--momentum tensor can be represented in the operator form~\cite{Adler}:
\beqn
{T^\alpha_\alpha(x)} & = & \frac{\beta(e)}{2 e}  F^{\mu\nu}(x) F_{\mu\nu}(x) + (1+\gamma_m) m \bar{\psi}\psi
\label{eq:Tmunu:avr_massive} \\
& & + \, {\mbox{discontinuous terms}}\,,
\nonumber
\eeqn
where $\gamma_m =  3\alpha_{\QED}/2\pi + \dots$ is the mass anomalous dimension, $\alpha_\QED = e^2/4\pi$ is the fine structure constant, and the ellipsis denote higher order  corrections in $\alpha_\QED$. It is however more illuminating to derive Eq.~\eq{eq:Tmunu:avr_massive} using the Fujikawa method in the path integral formalism~\cite{Fujikawa:1980vr} which attributes a leading $O(\alpha^1_{\QED})$ part of the first term in Eq.~\eq{eq:Tmunu:avr_massive} to the contribution from the integration measure over the fermionic fields. The remaining terms appear due to virtual photons and non-anomalous contributions originating from the classically non-conformal mass term of the action~\eq{eq:L:QEDm},
which include higher-order $O(\alpha_{\QED}^n)$ terms with $n \ge 1$.

The electric current generated by the scale magnetic effect depends on the expectation value of the trace of the energy-momentum tensor in the background of the classical gauge field $A_\mu^\cl$~\cite{Chernodub:2016lbo}. In the massive theory this trace includes both anomalous and non-anomalous parts coming from the quantum measure and the classical action, respectively~\eq{eq:Tmunu:avr_massive}. The anomalous part gives the known contribution to the current~\eq{eq:SME}. As for the non-anomalous part, we may only say that in the heavy mass limit ($m\to \infty$) the fermions should decouple from the dynamics of the model~\cite{Appelquist:1974tg} and therefore the contribution from the non-anomalous part should cancel the anomalous term~\eq{eq:SME}.

 Let us consider QED~\eq{eq:L:QEDm} in the classical electromagnetic field background with the field strength $F^\cl_{\mu\nu}$ in addition to the dynamical photons. Using symmetry properties as well as dimensional arguments one finds that the leading terms of the local derivative expansion in the dimensional regularization give $\langle (1+\gamma_m) m \bar{\psi}(x) \psi(x) \rangle = C_1 \, m^4 + C_2\, F^{\cl,\mu\nu}(x) F^\cl_{\mu\nu}(x) + O(m^{-2})$, where the constants $C_1$ and $C_2$ may, in principle,  contain a divergent dependence on the parameter $\epsilon = D-4$ of the dimensional regularization as well as a dependence on the dimensional parameter $\mu$ through the combination ${\rm log}\, (\mu/m)$.  The first term in this expression is irrelevant for the dynamics of the model. The factor in front of the field-dependent term is then fixed by the decoupling theorem:
\beqn
\left\langle (1+\gamma_m) m \bar{\psi}(x) \psi(x) \right\rangle & = & \mbox{const}  -  \frac{\beta(e)}{2 e} F^{\cl,\mu\nu}(x) F^\cl_{\mu\nu}(x) \nonumber \\
& & + O(m^{-2})\,.
\eeqn
This expression is valid in any regularization in the limit, when both the classical electromagnetic field and gradients of the metric are smaller than the corresponding power of the fermion mass $m$.

Equation~\eq{eq:Tmunu:avr_massive} is consistent with the existing calculations of the triangle correlator $\avr{TJJ}$ of the fermionic stress tensor $T$ and two external electric currents $J$ which may alternatively be used to compute the electric current generated by the scale electric and magnetic effects. The $\avr{TJJ}$ correlator vanishes in the limit when the large fermionic mass $m$ exceeds external momenta associated with the vertices of the triangle diagram~\cite{Giannotti:2008cv,Armillis:2009pq}, thus indicating that the induced electric current should also vanish in the $m \to \infty$ limit.  It was proposed that the dependence on the mass $m$ enters the anomalous relation~\eq{eq:Tmunu:avr} through the modified effective $\beta$ function~\cite{Armillis:2009pq}:
\beqn
\beta(p^2, m^2,M^2) = - e\, p^2 \frac{d \Pi_R(p^2, m^2,M^2)}{d p^2}\,,
\label{eq:beta:effective}
\eeqn
determined via the renormalized photon self-energy
\beqn
\Pi_R(p^2, m^2,M^2) = \Pi(p^2, m^2) - \Pi(p^2 = - M^2, m^2). \qquad
\label{eq:pi:R}
\eeqn
It is worth mentioning, that this definition of the beta function differs slightly from a standard textbook definition as the latter is determined by the dependence of the polarization operator on the mass scale $M$ rather than on the value of the momentum $p$.  The two definitions coincide in the ultraviolet limit $M \to \infty$. Therefore, the concrete form of Eq. (\ref{eq:beta:effective}) depends on particularities of the regularization scheme~\cite{Armillis:2009pq,Goncalves:2009sk}.

Notice that in an alternative heat kernel approach the correlator $\avr{TJJ}$ can be computed using different calculation schemes~\cite{Goncalves:2009sk} which give different expressions in the domain $p^2 \ll m^2$. However, all computations share the same qualitative feature: the correlator $\avr{TJJ}$ tends to zero in the infrared region $p^2 \ll m^2$~\cite{Giannotti:2008cv,Armillis:2009pq,Goncalves:2009sk} .

In the present paper we develop the Wigner-Weyl approach to compute the electric current generated by the scale magnetic effect. In agreement with the decoupling theorem, we explicitly demonstrate that the electric current of massive fermions contains both anomalous and non-anomalous  contributions which cancel each other exactly in the large mass limit.

\section{QED in curved spacetime}
\label{eq:curved}

\subsection{Fermionic lagrangian in curved spacetime}

A Dirac fermion field $\psi$ with the mass $m$ in a curved background is described by the action
\beqn
S = \int d^4 x \sqrt{- g} \cL\,,
\label{eq:S}
\eeqn
with the following Lagrangian~\cite{ref:book}:
\beqn
\cL = \frac{i}{2} \left[ \bar\psi e^{\mu}_{\;a} \gamma^a \nabla_\mu \psi - \left( \nabla_\mu \bar\psi \right) e^{\mu}_{\;a}\gamma^a \psi \right] - m \bar\psi\psi\,,
\label{eq:L}
\eeqn
where $\nabla_\mu \bar\psi \equiv (\nabla_\mu \psi)^\dagger \gamma^0$ and $\gamma^a$ are the standard, coordinate-in\-de\-pen\-dent Dirac matrices. The vierbein (tetrad) field~$e^{\mu}_{\;a} \equiv e^{\mu}_{\;a}(x)$ is related to the spacetime metric $g_{\mu\nu}$ as follows: $g_{\mu\nu} = e_{\mu}^{\;a} e_{\nu}^{\;b} \eta_{ab}$, where $\eta_{ab}$ is the metric of the flat Minkowski spacetime and $e^{\mu}_{\;a} = g^{\mu\nu} \eta_{ab} e_{\nu}^{\;b}$. The raising/lowering of the curved spacetime indices (denoted by the Greek letters $\mu,\nu,\dots$) and of the flat indices (denoted by the Latin letters $a,b,\dots$) of the vierbein $e^{\mu}_{\;a}$ are done with respect to the curved metric $g_{\mu\nu}$ and the flat metric $\eta_{ab}$, respectively. For example, $e^{\mu a} = g^{\mu\nu} e_\nu^{\;a} = g^{\mu\nu} \eta^{ab} e_{\nu b}$, etc.

The covariant derivative
\beqn
\nabla_\mu = D_\mu +  \Gamma_\mu
\,, \qquad D_\mu = \partial_\mu + i e A_\mu\,,
\label{eq:nabla}
\eeqn
enforces the invariance of the Lagrangian~\eq{eq:L} with respect to the U(1) gauge transformations
\beqn
\begin{array}{rcl}
\psi(x) & \to & e^{i \alpha(x)} \psi(x)\,,
\quad
\bar\psi(x) \to e^{-i \alpha(x)} \bar\psi(x)\,, \\[3mm]
A_\mu(x) & \to & A_\mu(x) - \frac{1}{e} \partial_\mu \alpha(x)\,,
\end{array}
\label{eq:U1:gauge}
\eeqn
and local Lorentz transformations in the curved spacetime $x^\mu \to x^{\mu'} = \Lambda^\mu_{\;\nu} x^\nu$. The latter is done with the help of the (matrix) spin connection
\beqn
\Gamma_\mu = - \frac{i}{4} \omega^{\;ab}_\mu \sigma_{ab}\,,
\label{eq:Gamma:mu}
\eeqn
where
\beqn
\sigma_{ab} = \frac{i}{2} \left[ \gamma_a,\gamma_b \right]
\eeqn
is the generator of the Lorentz transformations and
\beqn
\omega^{\;ab}_\mu = e_\nu^{\;a} \Gamma^\nu_{\;\sigma\mu} e^{\sigma b} + e_\nu^{\;a} \partial_\mu e^{\nu b}\,,
\eeqn
with the Christoffel symbols
\beqn
\Gamma^\nu_{\;\alpha\mu} = \frac{1}{2} g^{\nu \beta} \left( \partial_\alpha g_{\beta \mu} +  \partial_\mu g_{\alpha\beta} -  \partial_\beta g_{\alpha \mu}\right).
\eeqn
A variation of the Lagrangian~\eq{eq:L} with respect to ${\bar\psi}$ leads to the covariant Dirac equation:
\beqn
\left( i e^{\mu}_{\;a} \gamma^a \nabla_\mu - m \right) \psi = 0\,.
\eeqn

Below we consider the conformally flat metric~\eq{eq:g:munu}. The corresponding components of the vierbein can be chosen as follows: $e^{\;a}_\mu = e^{\tau} \delta^{a}_\mu$ so that $e^{\mu a} = e^{- \tau} \eta^{\mu a}$ and
\beqn
e^{\mu}_{\;a} = e^{-\tau} \delta_{a}^\mu\,.
\label{eq:e}
\eeqn
The metric determinant in~\eq{eq:S} is $g \equiv \det g_{\mu\nu} = - e^{8\tau}$.

In the conformally flat metric~\eq{eq:g:munu} the spin connection part~\eq{eq:Gamma:mu} in the Lagrangian~\eq{eq:L} takes the following form:
\beqn
e^\mu_{\;a} \gamma^a \Gamma_\mu = {\gamma^a} {\mathit \Gamma}_a(x),
\qquad
{\mathit \Gamma}_a = \frac{3}{2} e^{-\tau} \partial_a \tau, \qquad
\label{eq:Gamma:mu:flat}
\eeqn
where we took into account that
$\omega^{ab}_\mu = \delta^a_\mu \partial^b \tau -\delta^b_\mu \partial^a \tau$.

Generally, the curved background affects the fermionic fields via the volume element $\sqrt{-g}$ in the action~\eq{eq:S}, the vierbein field~$e^{\mu}_{\;a}$ and the spin connection ${\mathit \Gamma}_\mu$ in the Lagrangian~\eq{eq:L} and \eq{eq:nabla}. However, in the conformal background~\eq{eq:g:munu} the contribution from the spin connection drops out from the Lagrangian~\eq{eq:L} because the spin connection~\eq{eq:Gamma:mu:flat} is a real-valued vector  ${\mathit \Gamma}_\mu^* \equiv {\mathit \Gamma}_\mu$ proportional to the identity matrix in the spinor space. Then the Lagrangian~\eq{eq:L:2} gets simplified:
\beqn
\cL = \frac{i}{2} \left[ \bar\psi e^{\mu}_{\;a} \gamma^a D_\mu \psi - \left( D_\mu \bar\psi \right) e^{\mu}_{\;a} \gamma^a \psi \right] - m \bar\psi\psi\,,
\label{eq:L:2}
\eeqn
where the electromagnetic covariant derivative $D_\mu$ is given in Eq.~\eq{eq:nabla}.

\subsection{Partition function and electric current}

We consider fermions in the fixed curved spacetime given by the metric $g_{\mu\nu}$ subjected to the fixed background of external electromagnetic field $A_\mu$. The fermionic partition function
\beqn
\cZ[A,g] & {=} & {\int} D {\bar\psi} D \psi \, e^{i S} \nonumber \\
& \equiv & {\int} D {\bar\psi} D \psi \exp\Bigl\{ i
{\int} d^4 x\, {\bar \psi}(x) \cD[A,g,m] \psi (x) \Bigr\} \nonumber \\
& = & {\mathrm{const}}\cdot\det \cD[A,g,m]\,,
\eeqn
is proportional to the determinant of the fermionic operator $\cD$ which enters the Lagrangian density~\eq{eq:L} in the action~\eq{eq:S}:
\beqn
\sqrt{-g(x)} \cL(x) = {\bar \psi}(x) \cD[A,g,m] \psi (x)\,.
\label{eq:D:definition}
\eeqn

In this article we consider the scale magnetic effect which arise in thermal equilibrium in static gravitational field with a time-independent metric~\eq{eq:g:munu} in background of external magnetic field ${\bs B} \neq 0$. The electric field is zero ${\bs E} = 0$. Since we consider the system in thermal equilibrium it is convenient to make the Wick rotation of the time coordinate $x_0 \to x_4 = - i x_0$ and formulate the theory in the Euclidean four-dimensional spacetime. The operator $\cD$ in the Euclidean space can explicitly be calculated with the help of Eqs.~\eq{eq:nabla}, \eq{eq:e}, \eq{eq:L:2} and \eq{eq:D:definition}:
\beqn
\cD & = & - \frac{i}{2} \sum_{\mu=1}^4 \gamma_\mu \Bigl[e^{3 \tau(x)} \frac{\partial}{\partial x^\mu} + \frac{\partial}{\partial x^\mu}  e^{3 \tau(x)} \Bigr]
\nonumber \\
& & - e^{3 \tau(x)}  \sum_{\mu=1}^4 \gamma_\mu e A_\mu(x) - i e^{4 \tau(x)} m, \qquad
\label{eq:cD:E}
\eeqn
where $\gamma_\mu$ are the Euclidean Dirac matrices. Correspondingly, we have
\beqn
\cZ[A,g] & {=} & {\int} D {\bar\psi} D \psi \exp\Bigl\{ - {\int} d^4 x\, {\bar \psi}(x) \cD[A,g,m] \psi (x) \Bigr\}. \qquad
\label{eq:Z:Euclidean}
\eeqn

The local electric current induced by the external gauge field $A_\mu$ in the curved spacetime background $g_{\mu\nu}$ is given by the following variational derivative:
\beqn
J^\mu(x;A,g) =  -\frac{\delta \log \cZ[A,g]}{\delta A_\mu(x)}\,.
\label{eq:J:one}
\eeqn
In the flat space with the Euclidean metric
$\eta_{\mu\nu} = \delta_{\mu\nu}$
the integration measure over the fermion field $D_\eta \bar{\psi} D_\eta \psi$ in \eq{eq:Z:Euclidean} is independent of the gauge field. In the curved background with the conformal metric~\eq{eq:g:munu} the integration measure $D_g \bar{\psi} D_g \psi $ acquires a dependence on the external gauge field \cite{Fujikawa:1980vr}:
\beqn
& & D_g \bar{\psi} D_g\psi =  D_\eta \bar{\psi}^\tau D_\eta \psi^\tau  \nonumber\\
& & \qquad \qquad \cdot \exp\Big\{\frac{\beta^{\mathrm{1loop}}_{{\text{QED}}}}{2 e}\, \int d^4 x \, \tau(x)\, F^{\mu\nu}(x) F_{\mu\nu}(x)  \Big\},
\qquad
\label{measure}
\eeqn
where $\psi^\tau(x) = e^{\frac{3}{2}\tau(x)} \psi(x)$, while
\beqn
\beta^{\mathrm{1loop}}_{{\text{QED}}} = \frac{e^3}{12 \pi^2},
\label{eq:beta:QED}
\eeqn
is the one--loop QED beta function. Equation~(\ref{measure}) originates from the transformation of measure given in \cite{Fujikawa:1980vr} under Weyl transformations
\beqn
\psi(x) & \to & e^{-3\tau(x)/2}\psi(x),\\
g_{\mu\nu}(x) & \to & e^{2\tau(x)}g_{\mu\nu}(x)\,.
\eeqn

Since we consider the response of the virtual fermions on the background electromagnetic field in the vacuum, then the background field is assumed to be induced by an external electric current located outside of the considered region of space. In this case we have two contributions to the induced electric current
\beqn
J^\mu(x;A,g) = J^\mu_{\measure}(x;A,g) +J^\mu_{\action}(x;A,g), \quad
\eeqn
given, respectively, by the one-loop anomalous contribution from the fermionic integration measure
\beqn
J^\mu_{\measure}(x;A,g)
&\equiv &  - \frac{2\beta^{(1)}(e)}{ e}\,F^{\mu\nu}(x)   \partial_{\nu} \tau(x)\,,\nonumber\\ && + \frac{2\beta^{(1)}(e)}{ e}\,\tau(x) \partial_{\nu} F^{\mu\nu}(x)
\label{eq:J:0}
\eeqn
Here the second line is proportional to the external current that creates the given external field. We assume that it is localized outside the region of observations.
The remaining contribution comes from the classical action:
\beqn
J_{\rm action}^\mu(x;A,g) =  -\frac{\delta \log \cZ_\eta[A,g]}{\delta A_\mu(x)}\,.
\label{eq:J:one2}
\eeqn
where
\beqn
\cZ_\eta[A,g]  & {=} &  {\int} D_\eta {\bar\psi^\tau} D_\eta \psi^\tau e^{ - {\int} d^4 x\, {\bar \psi}(x) \cD[A,g,m] \psi (x) }
\label{eq:Z:Euclidean2}\\
& {=} & {\int} D_\eta {\bar\psi}^\tau D_\eta \psi^\tau e^{ - {\int} d^4 x\, {\bar \psi}^\tau(x) \cD[e^{\tau}A,\eta,e^{\tau}m] \psi^\tau (x)}.
\nonumber
\eeqn

\subsection{The case of massless fermions}

One can see, that even for \revision{a} vanishing mass there may be an extra contribution to \revision{the} electric current given by
\beqn
J^\mu_{\action}(x;A,g)\Big|_{m=0}
\equiv e^{\tau(x)}\Tr \biggl[G_\eta \frac{\delta \cD[A^\tau,\eta,0]}{\delta A^\tau_\mu(x)} \biggr]\,. \quad
\label{eq:J:01}
\eeqn
where $A^\tau(x) = e^{\tau(x)}A(x)$ \revision{and the Green function}
\revisionZZ{\beqn
G_\eta(x,y) & {=} & \frac{1}{Z_\eta[A,g]} {\int} D_\eta {\bar\psi}^\tau D_\eta \psi^\tau \, \psi^\tau(x) {\bar \psi}^\tau(y)
\label{eq:G:def} \\
&& \exp\Bigl\{ - {\int} d^4 x\, {\bar \psi}^\tau(x)\cD[e^\tau A,\eta,0] \psi^\tau (x) \Bigr\}, \quad
\nonumber\eeqn}
satisfies \revision{the} relation:
\beqn
\cD[ A^\tau,\eta,0] G_\eta(x,y) = \delta^{(4)}(x-y) \,.
\label{eq:G:eq}
\eeqn
The field $A^{\tau}(x)$ gives rise to \revision{the} ``magnetic'' field $\partial_{[i} A^\tau_{j]} \epsilon^{ijk0}$ and \revision{to the} "electric" field $\partial_{[0} A^\tau_{k]} $. According to the results of \cite{Zubkov:2016mvv} in such systems the vacuum current proportional to the first power of "magnetic" or "electric" field is \revision{also} proportional to the topological invariant in momentum space, which is vanishing for the system under consideration. The terms \revision{linear in} the first derivatives of ``magnetic'' or ``electric'' field might appear with dimensionless coefficient. If exists, such a term would have the form $J^k_{{\rm action}}(x) = {\rm const}\, e^{\tau(x)} \partial_i \partial^{[i} A^{\tau,k]}(x)$, i.e. it should be proportional to the electric current that creates the given external field. \revision{Essentially, it} is the renormalization of this current due to the quantum fluctuations as well as \revision{due to contribution from} the second line in Eq. (\ref{eq:J:0}). In our consideration we assume that such a current is localized far outside the region of observations. This \revision{assumption} ensures that in \revision{the relevant} order the extra contribution to the SME current is absent. The terms proportional to the second power of ``magnetic'' field in \revision{the conformal limit}
are to be suppressed as $1/\Lambda$, where $\Lambda$ is the ultraviolet cutoff. The same \revision{conclusion} is valid for the higher order corrections. Overall, the component $J^k_{{\rm action}}$ vanishes to all orders for the system of massless fermions.

\subsection{The general case}

Below we consider the case of massive fermions, where \revision{the current} $J^k_{{\rm action}}$ remains nonvanishing. We will explore \revision{the following} relation
\beqn
J^\mu_{\action}(x;A,g)
\equiv \Tr \biggl[G \frac{\delta \cD[A,g,m]}{\delta A_\mu(x)} \biggr]\,,
\label{eq:J:1}
\eeqn
\revision{where}\revisionZZ{
\beqn
& & G(x,y) = \frac{1}{Z_\eta[A,g]}{\int} D_\eta {\bar\psi}^\tau D_\eta \psi^\tau \, e^{-3(\tau(x)+\tau(y))/2} \nonumber\\&&
\psi^\tau(x) {\bar \psi}^\tau(y) e^{ - {\int} d^4 x\, {\bar \psi}^\tau(x)e^{-3\tau(x)/2} \cD[A,g] e^{-3\tau(x)/2} \psi^\tau (x) }, \quad\
\label{eq:G:def}
\eeqn}
is the fermionic Green function satisfying the relation:
\beqn
e^{3(\tau(y)-\tau(x))/2}\cD[A,g] G(x,y) = \delta^{(4)}(x-y) \,,
\label{eq:G:eq}
\eeqn
\revision{which is equivalent to the following equation:}
\beqn
\cD[A,g] G(x,y) = \delta^{(4)}(x-y) \,.
\label{eq:G:eq}
\eeqn
According to Eqs.~\eq{eq:nabla}, \eq{eq:e} and \eq{eq:L:2} the variation of the local operator $\cD$ with respect to the gauge potential $A_\mu$ in Eq.~\eq{eq:J:1} gives the following ultra-local two-point operator:
\beqn
\biggl(\frac{\delta \cD}{\delta A_\mu(x)}\biggr)(y,z) & = & -\sqrt{-g(x)} e^\mu_{\;a}(x) \gamma^a \bbbone_{x,y} \bbbone_{x,z} \nonumber \\
& = &- e\, e^{- 3 \tau(x)} \gamma^\mu \bbbone_{x,y} \bbbone_{x,z} \,,
\label{eq:D:D:A}
\eeqn
where we denoted for compactness
\beqn
\bbbone_{x,y} = \delta^{(4)}(x-y)\,.
\label{eq:unity}
\eeqn
The first prefactor ``$e$'' in the last line of Eq.~\eq{eq:D:D:A} is the electric charge. It is then convenient to rewrite the induced electric current~\eq{eq:J:1} in the following compact form:
\beqn
J^\mu(x) = -e\, e^{- 3 \tau(x)} \Tr_{\!y,z} \bigl[G(y,z) \cdot \gamma^\mu \bbbone_{x,y} \bbbone_{x,z} \bigr], \qquad
\label{eq:J:2}
\eeqn
where the exponential prefactor corresponds to a trivial conformal volume factor coming from the fact the electric current has the dimension $[\mathrm{mass}]^3$.

Technically, our aim is to calculate the electric current~\eq{eq:J:2} with the help of the Green function~\eq{eq:G:def} determined by Eqs.~\eq{eq:G:eq} and \eq{eq:cD:E}. To this end we will use the Wigner-Weyl formalism described in the next section.

\section{Wigner-Weyl formalism}
\label{ref:Wigner}

Let us very briefly review basic features of Wigner functions and Weyl symbols in quantum mechanics that we will later use in the quantum field theory. A pedagogical overview of the Wigner-Weyl quantization formalism may be found, for example, in reviews~\cite{ref:pedestrians,ref:Polkovnikov,star}. We choose the system of units $\hbar = c = 1$ and work in the $(3+1)$ dimensional spacetime.

Let $\hat A$ be an operator which is a function of the position operator ${\bs {\hat x}}$ and the momentum operator ${\bs{\hat p}}$ which obey the standard commutation rule:
\beqn
[{\hat x}^k, {\hat p}^l] = i \delta^{kl}\,.
\eeqn
The Weyl symbol $\widetilde A$ of the operator $\hat A$ is a function of the three-dimensional coordinate $\bs x$ and momentum $\bs p$ which is given by the following Wigner transformation~\cite{Wigner:1932eb,berezin,Littlejohn:1985ht}:
\beqn
{\widetilde A}({\bs x}, {\bs p}) {=} \int d^3 r e^{- i {\bs p} {\bs r}} \bra{{\bs x}  - {\bs r}/2} {\hat A}({\hat {\bs x}},{\hat {\bs p}}) \ket{{\bs x} + {\bs r}/2}, \qquad
\label{eq:Wigner:transform}
\eeqn
which is expressed via matrix elements $\bra{{\bs x}}{\hat A}\ket{{\bs x}'}$ of the operator ${\hat A}$ in the basis of wavefunctions $\ket{{\bs x}}$ labelled by the coordinate ${\bs x}$. The Wigner transformations maps operators to functions.

The Wigner function\footnote{We rescale the Wigner function in Eq.~\eq{eq:W:0} by the factor $(2 \pi)^3$ compared to a standard de\-fi\-ni\-tion~\cite{ref:pedestrians} in order to keep a conventional form of the phase-space volume in Eq.~\eq{eq:average} and thereafter.}
\beqn
W({\bs x}, {\bs p}) = \int d^3 r e^{- i {\bs p} {\bs r}} \bra{{\bs x} - {\bs r}/2} {\hat \rho} \ket{{\bs x} + {\bs r}/2}\,, \qquad
\label{eq:W:0}
\eeqn
is the Wigner transform~\eq{eq:Wigner:transform} of the density matrix operator~$\hat \rho$. For pure states ${\hat \rho} = \ket{\psi} \bra{\psi}$, the Wigner function~\eq{eq:W:0}   can be directly expressed via the wavefunctions $\psi({\bs x}) = \avr{{\bs x} | \psi}$ as follows:
\beqn
W({\bs x}, {\bs p}) = \int d^3 r e^{- i {\bs p} {\bs r}} \psi({\bs x} - {\bs r}/2) \psi^*({\bs x} + {\bs r}/2)\,, \qquad
\label{eq:W:1}
\eeqn

The Wigner-Weyl formalism has many useful features. A trace of two operators $\hat A$ and $\hat B$ is given by a convolution of their Weyl symbols over the whole phase space:
\beqn
\Tr \bigl( {\hat A} {\hat B} \bigr) = \int \frac{d^3 x \, d^3 p}{(2 \pi)^3} {\widetilde A}({\bs x}, {\bs p}) {\widetilde B}({\bs x}, {\bs p})\,.
\label{eq:trace}
\eeqn
Therefore, the expectation value of an operator ${\hat A}$ can be expressed as a convolution of the Weyl symbol of the operator ${\hat A}$ and the Wigner function $W$:
\beqn
\langle{{\hat A}}\rangle \equiv \Tr ({\hat \rho} {\hat A} ) = \int \frac{d^3 x \, d^3 p}{(2 \pi)^3} W({\bs x}, {\bs p}) {\widetilde A}({\bs x}, {\bs p})\,.
\label{eq:average}
\eeqn

Weyl symbols of certain operators are easy to calculate. For the purposes that will become evident below, let us consider the following operator
\beqn
{\hat K} ({\bs {\hat x}}, {\bs {\hat p}}) = {\hat A}({\bs {\hat p}}) + \frac{1}{2} \left[ B({\bs {\hat x}}) {\bs b} {\bs{\hat p}} + {\bs b} {\bs{\hat p}} B({\bs {\hat x}}) \right]
+ {\hat C}({\bs {\hat x}}), \quad
\label{eq:separable:hat:K}
\eeqn
where ${\bs b}$ is a fixed vector, the operator ${\hat A}$ is a function of a momentum operator ${\bs{\hat p}}$ only while the operators  ${\hat B}$ and ${\hat C}$ depend only the coordinate operator ${\bs{\hat x}}$. The Weyl symbol of the operator~\eq{eq:separable:hat:K} is given by the sun of the corresponding functions:
\beqn
{\widetilde K} ({\bs { x}}, {\bs { p}}) = { A}({\bs { p}}) +
B({\bs { x}}) {\bs b} {\bs{ p}} + { C}({\bs { x}}), \quad
\label{eq:separable:tilde:K}
\eeqn
so that the Weyl transformation~\eq{eq:Wigner:transform} for the particular form of the operator~\eq{eq:separable:hat:K} amounts to the simple substitution $\bs {\hat x} \to \bs {x}$ and $\bs {\hat p} \to \bs {p}$. For more complex operators this is not the case.

The Weyl transform of a product of two operators $\hat D$ and $\hat G$ can be expressed in terms of the Wigner transformations of these operators via the Groenewold formula~\cite{ref:Groenewold}:
\beqn
\widetilde{DG}({\bs x}, {\bs p}) = {\widetilde D}({\bs x}, {\bs p}) \star {\widetilde G}({\bs x}, {\bs p})\,,
\label{eq:Groenewold}
\eeqn
where the Moyal (star) product~\cite{ref:Moyal}
\beqn
\star \equiv e^{\frac{i}{2} \!\overset{\leftrightarrow}{\partial}_{\!{\bs {xp}}}} = 1 + \frac{i}{2} \overset{\leftrightarrow}{\partial}_{\!{\bs {xp}}}
- \frac{1}{8} \overset{\leftrightarrow}{\partial^{2}}_{\!\!\!{\bs {xp}}} + \dots\,,
\label{eq:star:expansion}
\eeqn
is essentially an exponentiation of the Poisson bracket kernel which can be expanded in powers of the double-derivative operator~\cite{Littlejohn:1985ht}:
\beqn
\overset{\leftrightarrow}{\partial}_{\!{\bs {xp}}}
= \overset{\leftarrow}{\partial_{\bs x}} \overset{\rightarrow}{\partial_{\bs p}}
- \overset{\leftarrow}{\partial_{\bs p}} \overset{\rightarrow}{\partial_{\bs x}}\,,
\label{eq:Janus}
\eeqn
 that acts both on the left and right sides (for example, $f\overset{\leftrightarrow}{\partial}_{\!{\bs {xp}}}g =
\partial_{\bs x} f \partial_{\bs p} g - \partial_{\bs p} f \partial_{\bs x} g$, etc).

Stationary systems may be described by the time-independent Wigner function~\eq{eq:W:1}. Non-stationary processes may be treated with the help of the time-dependent Wigner function $W({\bs x}, {\bs p};t)$ in which the time variable enters in a different way compared to the spatial coordinates. The time evolution of the Wigner function is determined by the Hamiltonian of the system $H$ via the Moyal (star) bracket~\eq{eq:Janus}~\cite{ref:Moyal}:
\beqn
i \frac{\partial }{\partial t} W({\bs x}, {\bs p};t) = H \star W({\bs x}, {\bs p};t) - W({\bs x}, {\bs p};t) \star H. \quad
\eeqn

Before going into further details we would like to highlight why the Wigner approach is a particularly useful method for our problem. We calculate a quantum average~\eq{eq:J:2} of a current $j$ which is given by a trace of the product $\avr{j} = \Tr [G B]$ of a Green function $G$ and a simple operator $B$. The trace can be calculated as the convolution~\eq{eq:trace} of the corresponding Weyl symbols $\widetilde G$ and $\widetilde B$. We will see that the Weyl symbol $\widetilde B$ may be easily obtained by the Weyl transformation~\eq{eq:Wigner:transform} of the operator $B$ itself, while the Weyl symbol $\widetilde G$ of the Green function $G$ may be calculated using the Groenewold formula~\eq{eq:Groenewold} applied to the identity $\bbbone = D G$, where $D$ is an operator which possesses the Weyl symbol $\widetilde D$ of a simple functional form. Since ${\widetilde \bbbone} \equiv 1$, the Groenewold equation~\eq{eq:Groenewold} transforms to $1 = {\widetilde D} \star {\widetilde G}$ which can be solved iteratively with respect to the Green function ${\widetilde G}$ in terms of the gradient expansion~\eq{eq:star:expansion} of the exponentiated double-derivative operator~\eq{eq:Janus}. This strategy -- which has been applied first time to Euclidean quantum field theory in Ref.~\cite{Zubkov:2016mvv} and is common for non-commutative field theories~\cite{Szabo:2001kg} -- will be realized in the next section.

\section{The non-anomalous contribution to the electric current}
\label{ref:current}

\subsection{Closed form of electric current}

In order to determine the non-anomalous contribution $J_{\action}$ to the generated electric current we apply the Wigner-Weyl formalism to the vacuum of QED in nontrivial gravitational and electromagnetic backgrounds. We are interested in stationary effects in thermal equilibrium in three spatial dimensions ${\bs x} = (x_1,x_2, x_3)$ which may be formulated in Euclidean four-dimensional space in which the fourth ``time'' coordinate plays a role of the imaginary time $x_4$. In order to ensure the validity of the Wick rotation of our Euclidean results back to Minkowski spacetime we assume that the electric field is vanishing (it would otherwise be imaginary in the Euclidean space) and that the metric is a time-independent quantity.

The Wigner-Weyl formalism may naturally be generalized to the four-dimensional Euclidean space in which the four-component coordinate operator ${\hat x} = ({\hat x}_1, \dots ,{\hat x}_4)$ is conjugated with the four-component momentum operator~${\hat p}$. This technique, which utilizes the Groenewold formula~\eq{eq:Groenewold} and the derivative expansion of the Moyal product~\eq{eq:star:expansion} at the level of the Green functions and the Weyl symbols of the corresponding operators, has been worked out in details in Ref.~\cite{Zubkov:2016mvv}. Below we describe essential details of the approach.

In the coordinate space momentum operator ${\hat p}$ takes the familiar form of the derivative operator ${\hat p}_\mu = - i \partial_{x_\mu}$ with $\mu = 1,\dots,4$ and the fermionic operator~\eq{eq:cD:E} takes the following form:
\beqn
{\hat \cD}({\hat x},{\hat p}) & = & \frac{1}{2} \Bigl[e^{3 \tau({\hat x})} {\hat {\slashed p}} + {\hat {\slashed p}} \, e^{3 \tau({\hat x})} \Bigr]
\nonumber \\
& & - e \slashed{A}({\hat x}) - i e^{4 \tau({\hat x})} m, \qquad
\label{eq:cD:E2}
\eeqn
where we used the standard ``slashed'' notation $\slashed{a} = \sum_{\mu=1}^4 \gamma_\mu a_\mu$.

The Weyl symbol of the fermionic operator ${\hat \cD}$ is given by the Wigner transformation~\eq{eq:Wigner:transform}:
\beqn
{\widetilde \cD}(x,p) & = & e^{3 \tau(x)} \bigl[{\slashed p} - e \slashed{A}(x)\bigr] - i e^{4 \tau(x)} m,
\label{eq:cD:W}
\eeqn
where we used the fact that the operator~\eq{eq:cD:E2} matches the general form of the operator~\eq{eq:separable:hat:K} with the known Weyl symbol~\eq{eq:separable:tilde:K}. Then the Weyl symbol for the fermionic Green function~\eq{eq:G:def} is formally given by the Wigner transform~\eq{eq:Wigner:transform}:
\beqn
{\widetilde G}(x,p) = \int d^4 r \, e^{-i p r} \, G(x - r/2,x + r/2)\,.
\label{eq:G:W:def}
\eeqn
This expression assumes that the Green function $G$ is defined in a background of a classical U(1) gauge field $A_\mu$ in a fixed gauge. Therefore one can suggest that Eq.~\eq{eq:G:W:def} does not require explicit introduction of a parallel gauge transport in a form of the Schwinger line $P(x,y)$ which is  an exponentiated gauge field integrated along an open contour joining the two points of the Green function~\eq{eq:G:W:def}. Below we explicitly demonstrate that the inclusion of the Schwinger line does not affect the final result for the electric current in the magnetic-field background. The presence of the Schwinger line is more suitable for systems with dynamical gauge fields~\cite{Vasak:1987um}.

Notice that in addition to the gauge transport line one could also expect the appearance of the spin connection transport. However, in the background of the conformally flat metric~\eq{eq:g:munu} the spin connection term does not enter the Lagrangian~\eq{eq:L:2} and therefore the parallel spin transport is trivial.

The Weyl symbol of the Green function~\eq{eq:G:W:def} can be calculated explicitly with the help of the Groenewold formula~\eq{eq:Groenewold} applied to Eq.~\eq{eq:G:eq}:
\beqn
1 = {\widetilde \cD}(x,p) \star {\widetilde G}(x,p)\,,
\label{eq:DG:W}
\eeqn
with ${\widetilde \cD}$ given explicitly in Eq.~\eq{eq:cD:W}. The star product in Eq.~\eq{eq:DG:W} is a straightforward generalization of Eqs.~\eq{eq:star:expansion} and \eq{eq:Janus} to the four-dimensional Euclidean space:
\beqn
\star \equiv e^{\frac{i}{2} \!\overset{\leftrightarrow}{\partial}} = 1 + \frac{i}{2} \overset{\leftrightarrow}{\partial}
- \frac{1}{8} \overset{\leftrightarrow}{\partial} {}^{2} + \dots\,,
\label{eq:star:4d}
\eeqn
with
\beqn
\overset{\leftrightarrow}{\partial} \equiv \overset{\leftrightarrow}{\partial}_{\!{{xp}}} = \sum_{\mu=1}^4 \left(
\overset{\leftarrow}{\partial}_{x_\mu} \overset{\rightarrow}{\partial}_{p_\mu}
- \overset{\leftarrow}{\partial}_{p_\mu} \overset{\rightarrow}{\partial}_{x_\mu}\right).
\label{eq:Janus:4d}
\eeqn

The non-anomalous electric current~\eq{eq:J:2} can be calculated with the help of a 4d generalization of the convolution formula~\eq{eq:trace}:
\beqn
J_{\action}^\mu(x) & = & - e\, e^{- 3 \tau(x)} \nonumber \\
& & \int \frac{d^4 s \, d^4p}{(2\pi)^4} {\mathrm{tr}} \bigl[{\widetilde G}(s,p) {\widetilde \bbbone}_x(s,p) \gamma^\mu \bigr], \qquad
\label{eq:J:convolution:0}
\eeqn
where the trace goes over the spinor indices only. The Wigner transform ${\widetilde \bbbone}_x(s,p)$ of the product ${\bbbone}_x(y,z) \equiv {\bbbone_{x,y} \bbbone_{x,z}}$ of the unit operators~\eq{eq:unity} can be calculated straightforwardly with the help of Eq.~\eq{eq:Wigner:transform}:
\beqn
{\widetilde{\bbbone}_x}(s,p) = \delta^{(4)}(x-s)\,,
\label{eq:unity:explicit}
\eeqn
where $s$ is the 4d spacetime coordinate and $p$ is the 4d momentum. Substituting Eq.~\eq{eq:unity:explicit} into Eq.~\eq{eq:J:convolution:0} we get the compact expression for the non-anomalous electric current via the Wigner transform of the fermionic propagator ${\widetilde G}(x,p)$:
\beqn
J_{\action}^\mu(x) = -e\, e^{- 3 \tau(x)} \int \frac{d^4p}{(2\pi)^4} {\mathrm{tr}} \bigl[{\widetilde G}(x,p) \gamma^\mu \bigr]. \qquad
\label{eq:J:convolution:1}
\eeqn

Now let us briefly demonstrate that the inclusion of the parallel gauge transport
\beqn
P(x,y) = \exp\left[ i e \int_x^y d x_\mu \, A^\mu(x)\right],
\label{eq:P}
\eeqn
in the definition of the Weyl symbol~\eq{eq:G:W:def} for the fermionic Green function~\eq{eq:G:def} gives us the gauge-invariant symbol
\beqn
{\widetilde G}_{\mathrm{inv}}(x,p) = \int d^4 r \, e^{-i p r} \, && P\left(x - r/2,x + r/2\right) \nonumber \\
& & G\left(x - r/2,x + r/2\right)
\label{eq:G:W:def:Schwinger}
\eeqn
which does not affect the result for the anomalous current~\eq{eq:J:convolution:1}. To this end we choose the contour connecting the points $x-r/2$ and $x+r/2$ in the form of a straight line
\beqn
{\bar x}^\mu(t) = x^\mu + \left(t - \frac{1}{2}\right) r^\mu\,,
\label{eq:line}
\eeqn
parameterized by the parameter $t \in [0,1]$. Taking the gauge potential of the magnetic field $B$ in the symmetric gauge, $A^\mu= (- e B x^2/2, e B x^1/2,0,0)$, we calculate the Schwinger line \eq{eq:P} with straight open contour~\eq{eq:line}, and then we get the following expression for the gauge-invariant Weyl-symbol~\eq{eq:G:W:def:Schwinger}:
\beqn
{\widetilde G}_{\mathrm{inv}}(x,p) & = & \int d^4 r \, e^{-i p r} \, e^{i (x_1 r_2 - x_2 r_1) eB/2} \nonumber \\
& & \hskip 18mm  G\left(x - r/2,x + r/2\right) \nonumber \\
& \equiv & {\widetilde G}(x,p-\bar{p}(x))\,,
\label{eq:G:W:def:Schwinger:2}
\eeqn
where the standard Weyl symbol~${\widetilde G}(x,p)$ is given in Eq.~\eq{eq:G:W:def} and $\bar{p}(x) = (B x^2/2, - B x^1/2,0,0)$. The next step is to calculate the anomalous current~\eq{eq:J:convolution:1} using the invariant Weyl symbol ${\widetilde G}_{\mathrm{inv}}$ of Eq.~\eq{eq:G:W:def:Schwinger:2} instead of the standard symbol ${\widetilde G}$. However, shifting the integration variable $p \to p + \bar{p}(x)$ we find that both definitions of the Weyl symbol~\eq{eq:G:W:def} and \eq{eq:G:W:def:Schwinger} lead to the same current of Eq. (\ref{eq:J:convolution:1}). Therefore, the Schwinger line (the parallel gauge transport) may indeed be ignored in the definition of the Weyl symbol in the background of the classical magnetic field.

Below we explicitly calculate the induced electric current~\eq{eq:J:convolution:1} in the leading order in derivative expansion of the star product~\eq{eq:star:4d}. The derivative series correspond also to a semiclassical expansion in the powers of the Planck constant $\hbar$. The latter is evident from the form of the exponential operator~\eq{eq:star:4d} in which the Planck constant is reinstated: $\star = \exp\{i \frac{\hbar}{2} \!\overset{\leftrightarrow}{\partial}\}$.

\subsection{Electric current in the leading order}

The electric current is given in Eq.~\eq{eq:J:convolution:1} where the Wigner transform of the fermionic propagator ${\widetilde G}(x,p)$ is completely determined by Eqs.~\eq{eq:cD:W}, \eq{eq:DG:W}, \eq{eq:star:4d} and \eq{eq:Janus:4d}. The Groenewold equation~\eq{eq:DG:W} for ${\widetilde G}$ can be solved iteratively in terms of the series:
\beqn
{\widetilde G} = {\widetilde G}^{(0)} +  {\widetilde G}^{(1)} +  {\widetilde G}^{(2)} + \dots\,.
\label{eq:tilde:G:expansion}
\eeqn
Notice that the iterative solution is a derivative expansion~\eq{eq:star:4d} as each power of the double-faced derivative~\eq{eq:Janus:4d} gives one power of a spatial derivative of either the conformal factor $\tau(x)$ or electromagnetic field $A_\mu(x)$.

The $n$-th order term ${\widetilde G}^{(n)}(x,p)$ is a local function of $x$ and $p$ proportional to the products of derivatives over $x$ of the form $(\partial^{l_1} \tau(x)) \dots (\partial^{l_L} \tau(x)) (\partial^{m_M} A(x)) \dots (\partial^{m_1} A(x))$ where the sum over the positive integers numbers $l_i$and $m_i$ equals to the order of the expansion, $l_1 + \dots + l_L + m_1 + \dots + m_M = n$. The electric current generated by the scale magnetic effect is given by the first-order (linear) response both in the conformal factor $\tau$ and in the electromagnetic field, $\partial_\alpha \tau \partial_\beta A_\gamma$, so that that the effect appears in the second-order term ${\widetilde G}^{(2)}$ in the expansion~\eq{eq:tilde:G:expansion}.

The zero-order term in the expansion~\eq{eq:tilde:G:expansion} is the usual (algebraic) inverse of the Weyl symbol~\eq{eq:cD:W} of the fermionic operator ${\hat \cD}$:
\beqn
{\widetilde G}^{(0)}(x,p) & = & {\widetilde \cD}^{-1}(x,p) \equiv 1/{\widetilde \cD}(x,p)
\label{eq:G:0}\\
& = & \frac{[{\slashed p} - e \slashed{A}(x)] + i e^{\tau(x)} m}{[p-eA(x)]^2 + e^{2 \tau(x)} m^2} e^{- 3 \tau(x)}\,.
\nonumber
\eeqn
By expanding the Groenewold equation~\eq{eq:DG:W} in powers of the double derivative $\dxp$, we express -- via the Weyl symbol~\eq{eq:cD:W} and its inverse~\eq{eq:G:0} -- the first order term:
\beqn
{\widetilde G}^{(1)} = - \frac{i}{2} {\widetilde \cD}^{-1}\! \left({\widetilde \cD}  \dxp {\widetilde \cD}^{-1}\!\right),
\eeqn
and then the second order term
\begin{subequations}
\beqn
{\widetilde G}^{(2)} & = & {\widetilde G}^{(2)}_I + {\widetilde G}^{(2)}_{II},
\label{eq:G2:sum}\\
{\widetilde G}^{(2)}_I & = & - \frac{1}{4} {\widetilde \cD}^{-1}\! \biggl\{
{\widetilde \cD}  \dxp\! \biggl[{\widetilde \cD}^{-1}\! \left({\widetilde \cD}  \dxp {\widetilde \cD}^{-1}\!\right)
\biggr]
\biggr\}, \\
{\widetilde G}^{(2)}_{II} & = & \frac{1}{8} {\widetilde \cD}^{-1}\! \left({\widetilde \cD}  \dxp{}^{2} {\widetilde \cD}^{-1}\!\right).
\eeqn
\label{eq:G2:1}
\end{subequations}

The second-order term~\eq{eq:G2:1} can further be rewritten as follows:
\begin{subequations}
\beqn
{\widetilde G}^{(2)}_{I} & {=} & - \frac{1}{4}
\bigl(R_\mu \partial_{p_\mu}-C_\mu \partial_{x_\mu} \bigr)\!\Bigl[\Bigr(C_\nu R_\nu - R_\nu C_\nu\Bigr) {\widetilde \cD}^{-1}\!\Bigr], \qquad
\label{eq:G2:I}\\
{\widetilde G}^{(2)}_{II} & {=} & \frac{1}{8} \Bigl[ R_{\mu\nu} \bigl(C_\mu C_\nu + C_\nu C_\mu - C_{\mu\nu} \bigr)  \nonumber \\
& & \ \ \; + C_{\mu\nu} \bigl(R_\mu R_\nu + R_\nu R_\mu - R_{\mu\nu} \bigr)
\label{eq:G2:II}\\
& & \ \ \; - S_{\mu\nu} \bigl(2 R_\mu C_\nu + 2 C_\nu R_\mu - S_{\mu\nu} - S_{\nu\mu} \bigr)
\Bigr] {\widetilde \cD}^{-1}.\nonumber
\eeqn
\label{eq:G2:2}
\end{subequations}
where
\begin{subequations}
\beqn
& & R_\mu = {\widetilde \cD}^{-1} \partial_{x_\mu} {\widetilde \cD}\,,
\qquad\quad\;\,
C_\mu = {\widetilde \cD}^{-1} \partial_{p_\mu} {\widetilde \cD}\,,
\label{eq:C:mu:2}\\
& & R_{\mu\nu} = {\widetilde \cD}^{-1} \partial_{x_\mu}\partial_{x_\nu} {\widetilde \cD}\,, \quad C_{\mu\nu} = {\widetilde \cD}^{-1} \partial_{p_\mu}\partial_{p_\nu} {\widetilde \cD}\,,
\label{eq:C:munu}\\
& & S_{\mu\nu} = {\widetilde \cD}^{-1} \partial_{p_\mu}\partial_{x_\nu} {\widetilde \cD}\,.
\eeqn
\label{eq:RC:definitions}
\end{subequations}
In deriving Eq.~\eq{eq:G2:2} we used the following identities:
\begin{subequations}
\beqn
& & \partial_{x_\mu} {\widetilde \cD}^{-1} = - R_\mu {\widetilde \cD}^{-1},
\quad\
\partial_{p_\mu} {\widetilde \cD}^{-1} = - C_\mu {\widetilde \cD}^{-1}, \\
& & \partial_{x_\mu} \partial_{x_\nu} {\widetilde \cD}^{-1} = \bigl(R_\mu R_\nu + R_\nu R_\mu - R_{\mu\nu} \bigr) {\widetilde \cD}^{-1}, \\
& & \partial_{p_\mu} \partial_{p_\nu} {\widetilde \cD}^{-1} = \bigl(C_\mu C_\nu + C_\nu C_\mu - C_{\mu\nu} \bigr) {\widetilde \cD}^{-1}, \\
& & \partial_{p_\mu} \partial_{x_\nu} {\widetilde \cD}^{-1} = \bigl(C_\mu R_\nu + R_\nu C_\mu - S_{\mu\nu} \bigr) {\widetilde \cD}^{-1}.
\eeqn
\end{subequations}
Notice that the quantities~\eq{eq:RC:definitions} are matrices in the spinor space and therefore in general they do not commute with each other. Moreover, $S_{\mu\nu} \neq S_{\nu\mu}$, while $R_{\mu\nu}  \equiv R_{\nu\mu}$ and $C_{\mu\nu}  \equiv C_{\nu\mu}$.

Using the relations
\begin{subequations}
\beqn
& & \partial_{x_\mu} R_\nu = R_{\mu \nu} - R_\mu R_\nu\,,
\quad
\partial_{p_\mu} C_\nu = C_{\mu \nu} - C_\mu R_\nu, \quad\\
& & \partial_{p_\mu} R_\nu = S_{\mu \nu} - C_\mu R_\nu\,,
\quad
\partial_{x_\mu} C_\nu = S_{\nu \mu} - R_\mu C_\nu,
\eeqn
\end{subequations}
we rewrite Eq.~\eq{eq:G2:I} as follows:
\beqn
& & {\widetilde G}^{(2)}_{I} = - \frac{1}{4} \Bigl[
   R_\nu \Bigl( C_{\mu\nu}  - \bigl\{ C_\mu, C_\nu \bigr\} \Bigr) R_\mu
\label{eq:G:2:II:2} \\
& & \qquad\qquad\ \,
- R_\nu \Bigl( S_{\nu\mu} - \bigl\{R_\mu, C_\nu \bigr\}\Bigr) C_\mu \nonumber \\
& & \qquad\qquad\ \,
+ C_\nu \Bigl( R_{\mu\nu} - \bigl\{R_\mu, R_\nu \bigr\}\Bigr) C_\mu \nonumber \\
& & \qquad\qquad\ \,
- C_\nu \Bigl( S_{\mu\nu} - \bigl\{C_\mu, R_\nu \bigr\}\Bigr) R_\mu \nonumber \\
& &
- R_\nu \bigl[C_\mu, R_\mu \bigr] C_\nu + C_\nu \bigl[C_\mu, R_\mu \bigr]  R_\nu \nonumber \\
& &
+ \bigl[R_\mu, C_\nu \bigr]  S_{\mu\nu} - R_\mu R_\nu C_{\mu\nu} - C_\mu C_\nu R_{\mu\nu}
\Bigr], \nonumber
\eeqn
where
\beqn
\bigl[A, B \bigr] = A B - B A,
\qquad
\bigl\{ A, B \bigr\} =  A B + B A,
\qquad
\label{eq:commutators}
\eeqn
are, respectively, the commutator and anticommutator.

The second-order correction ${\widetilde G}^{(2)}$ is now equal to the sum~\eq{eq:G2:sum} of the two terms ${\widetilde G}^{(2)}_I$ and ${\widetilde G}^{(2)}_{II}$ given, respectively, in Eqs.~\eq{eq:G2:II} and \eq{eq:G:2:II:2}. Since these terms do not contain external derivatives, we should expand them in the powers of the derivatives and keep eventually the terms containing the product $\partial_\mu \tau \partial_\nu A_\alpha$. As one can see from the explicit definitions~\eq{eq:cD:W} and \eq{eq:RC:definitions}, the relevant terms enter the $S$ and $R$ quantities:
\begin{subequations}
\beqn
&& S_{\mu\nu}(x,p) = 3 C_\mu(x,p) \partial_\nu \tau(x), \\
&& R_\mu(x,p) = \biggl[3 {-} \frac{i m}{{\widetilde \cD}(x,p)}\biggr] \partial_\mu \tau(x) {-} C_\nu(x,p) \partial_\mu A_\nu(x),\qquad  \\
&& R_{\mu\nu}(x,p) = - 3 \partial_{\{\mu,} \tau(x) \partial_{\nu\}} A_\alpha(x) C_\alpha(x,p).
\label{eq:Rmunu}
\eeqn
\label{eq:SRR}
\end{subequations}
In Eq.~\eq{eq:Rmunu} all irrelevant terms with double derivatives are not shown. The symmetrization with respect to the Lorentz indices is denoted by the curly brackets.

Equation~\eq{eq:SRR} indicates that every term in the second-order corrections to the Weyl symbol of the Green function, Eqs.~\eq{eq:G2:II} and \eq{eq:G:2:II:2} contains the required combination of the derivatives $\partial_\mu \tau \partial_\nu A_\alpha$. Since we are looking for the terms bilinear in $\tau$ and $A_\mu$, we keep the mentioned combination while setting $\tau$ and $A_\mu$ to zero in the prefactors of these terms. Denoting the latter with the superscript ``(0)'', one then immediately gets for the $S$ and $R$ Lorentz structures~\eq{eq:SRR} the following expressions:
\begin{subequations}
\beqn
&& S^{(0)}_{\mu\nu}(x,p) = 3 P_\mu(p) \partial_\nu \tau(x), \\
&& R^{(0)}_\mu(x,p) = \biggl[3 {-} i m P_0(p)\biggr] \partial_\mu \tau(x) {-} P_\nu(p) \partial_\mu A_\nu(x),\qquad  \\
&& R^{(0)}_{\mu\nu}(x,p) = - 3 \partial_{\{\mu,} \tau(x) \partial_{\nu\}} A_\alpha(x) P_\alpha(p),
\label{eq:Rmunu:2}
\eeqn
\label{eq:SRR:2}
\end{subequations}
where
\beqn
P_\mu(p) \equiv C^{(0)}_\mu(p) = \lim_{\tau \to 0} \lim_{A \to 0} C_\mu(x,p),
\label{eq:C:mu}
\eeqn
is the vector $C_\mu$ given by the second expression in Eq.~\eq{eq:C:mu:2} in flat space in the absence of external electromagnetic field.

Using the explicit form of the Weyl symbol~${\widetilde \cD}$ of the fermionic operator~\eq{eq:cD:W} and the second relation in Eq.~\eq{eq:C:mu:2} we explicitly get for Eq.~\eq{eq:C:mu}:
\beqn
P_\mu(p) = \frac{1}{p^2 + m^2}(\slashed p + i m) \gamma_\mu\,,
\label{eq:P:mu}
\eeqn
where $\gamma_\mu$ are Euclidean gamma matrices for $\mu=1,\dots,4$ and $\gamma_0 \equiv \bbbone$ is a unit matrix.\footnote{For convenience we complemented the four-vector~\eq{eq:P:mu} with the fifth $\mu =0$ component. In the Euclidean space our choice $\gamma_0 \equiv \bbbone$ does not interfere with the $\gamma_0$ matrix of the Minkowski space.} In addition, we notice that the Weyl symbol~${\widetilde \cD}$ is a linear function of the momentum~$p$, the second relation in Eq.~\eq{eq:C:munu} gives us $C_{\mu\nu} \equiv 0$.

Finally, we substitute the Lorenz structures~\eq{eq:Rmunu:2} and ~\eq{eq:C:mu} into the second-order corrections to the Weyl symbol of the Green function, Eqs.~\eq{eq:G2:II} and \eq{eq:G:2:II:2}, sum them up and put the result into the definition of the induced electric current~\eq{eq:J:convolution:1}. Then, in order keep the second-order corrections only, we set $\tau=0$ in the volume prefactor of the current~\eq{eq:J:convolution:1}, and after algebraic manipulations we
get the following compact expression for the generated non-anomalous electric current:
\beqn
J_{\action,\mu} =- \frac{i m e^2}{4} \frac{\partial \tau}{\partial x^\alpha} \frac{\partial A_\beta}{\partial x^\nu } \int \frac{d^4 p}{(2 \pi)^4} {\mathrm{tr}} \left( P_\mu P_{\nu\alpha\beta} \right), \qquad
\label{eq:j:5}
\eeqn
where the trace is taken over the spinor space while the tensor structure
\beqn
P_{\nu\alpha\beta} & = & \{ \{P_0,P_\nu\} , \{P_\alpha, P_\beta\} \} +  \{[P_0,P_\alpha], [P_\nu, P_\beta]\} \nonumber \\
& & - \{ \{P_0, P_\beta \} , \{P_\nu , P_\alpha \} \},
\label{eq:C:3}
\eeqn
is given in terms of the commutators and anticommutators~\eq{eq:commutators} of the matrices~\eq{eq:P:mu}.

The electric current~\eq{eq:j:5} is invariant under the U(1) gauge transformations~\eq{eq:U1:gauge} since the expression under the integral in Eq.~\eq{eq:j:5} is antisymmetric with respect to the interchange of the indices $\beta $ and $ \nu$. The latter fact can be checked directly by manipulation of Eq.~\eq{eq:C:3}. Therefore the derivative $\partial_\nu A_\beta$ in Eq.~\eq{eq:j:5} appears always in a form of the gauge-invariant electromagnetic field tensor $\partial_\nu A_\beta - \partial_\beta A_\nu \equiv F_{\nu\beta}$.

Substituting Eq.~\eq{eq:P:mu} into Eq.~\eq{eq:C:3} and taking the trace over the spinor indices we get the electric current~\eq{eq:j:5}:
\beqn
J_{\action,\mu} = - \alpha(m)  e^2 F_{\mu\nu} \partial_\nu \tau\,,
\label{eq:J:mu:1}
\eeqn
where the prefactor $\alpha(m)$ is given by the integral:
\beqn
\alpha(m) = 4 m^2 \int \frac{d^4 p}{(2 \pi)^4} \frac{(p^2 + 2 m^2)}{(p^2 + m^2)^4} \,,
\label{eq:C0}
\eeqn
which evaluates to a finite mass-independent quantity:
\beqn
\alpha(m) = \frac{1}{6 \pi^2}\,,
\label{eq:alpha:m}
\eeqn
We would like to remind that despite Eq.~\eq{eq:J:mu:1} has a visibly covariant 4-tensor form, our Euclidean derivation is formally valid only for the spatial indices $\mu$ and $\nu$ which do not allow us to consider either a nonzero background electric field, ${\bs E}$ or time-dependent conformal metric factor $\tau= \tau(t)$. Restricting ourselves to the case of the pure magnetic field background ${\bs B} \neq 0$ in spatially inhomogeneous curved space $\tau = \tau({\bs x})$ we obtain from Eqs.~\eq{eq:J:mu:1} and \eq{eq:alpha:m} the following non-anomalous contribution to the electric current in {\emph{massive}} QED:
\beqn
{\bs J}_{\action} = -\frac{e^2}{6 \pi^2} {\bs \nabla} \tau \times {\bs B}\,.
\label{eq:SME:2}
\eeqn
Taking into account the value of the one-loop QED beta function~\eq{eq:beta:QED}, we find that the part of the electric current~\eq{eq:SME:2} cancels precisely the one-loop anomalous part:
\beqn
{\bs J}_{\measure} & = &  \frac{2 \beta^{\mathrm{1loop}}_{{\text{QED}}}}{e} {\bs \nabla} \tau(x) \times {\bs B}(x)\,,
\label{eq:SME:measure}
\eeqn
which is coming from the integration measure~\eq{eq:J:0}. Therefore in the limit of heavy fermions the electric current generated by the SME vanishes:
\beqn
{\bs J} = {\bs J}_{\measure} + {\bs J}_{\action} = 0 + O\left(\partial^2/m^2\right),
\eeqn
where the second term denotes higher-derivative terms which are suppressed in the large-mass limit. These terms appear naturally in the derivative series~\eq{eq:tilde:G:expansion} which define iteratively the Wigner transform of the fermionic propagator as the solution of the Groenewold equation~\eq{eq:DG:W}.

\section{Discussions and Conclusion}

In our paper we discuss the scale magnetic effect (SME) which generates a vacuum electric current in external magnetic field in a curved spacetime~\cite{Chernodub:2016lbo}. The origin of the effect is the conformal anomaly which breaks, at the quantum level, a conformal symmetry in classically conformal gauge theories. This effect has already been considered in the QED with massless fermions. In our paper we ask the natural question what happens with the SME if the fermions are massive so that the conformal invariance is already explicitly broken at the classical level. In particular, we consider the limit when the fermion mass $m$ is much larger than the scale of external magnetic field $m^2 \gg |{\bs B}|$ and the scale of the spatial gradient the conformal factor $\tau$ of the metric, $m\gg |{\bs \nabla} \tau|$. This limit is opposite to the classically conformal case considered in Ref.~\cite{Chernodub:2016lbo}.

We demonstrate that the anomalous electric current generated by the massive fermions can conveniently be calculated by the Wigner-Weyl formalism which gives the derivative series inversely proportional to increasing powers of the fermion mass $m$. We have found that there are two contributions to the electric current generated by the SME.

The first contribution ${\bs J}_{\measure}$ comes from the integration measure over the fermionic fields. This current has the anomalous nature because the measure is not invariant under conformal (Weyl) transformations in the presence of the background electromagnetic field~\cite{Fujikawa:1980vr}. As a result, the anomalous contribution ${\bs J}_{\measure}$ does not depend on the fermion mass because the quantum measure is independent of the details of the classical fermionic action. At small fermion masses the anomalous term provides the major contribution to the SME current. The current ${\bs J}_{\measure}$ should also contain contributions induced by exchanges by virtual photons. These loop corrections are not considered in the present paper since they are suppressed by higher orders of the fine structure constant $\alpha_\QED$.

The second contribution ${\bs J}_{\action}$ originates from the classically nonvanishing terms in the trace of energy-momentum tensor. These terms appear due to the explicit breaking of the scale invariance at the level of the classical Lagrangian. Despite our derivation involved integrals in unbounded momentum space, the second contribution to electric current is a finite quantity both in ultraviolet and infrared regimes. The absence of the ultraviolet divergencies implies that the anomalous current does not require regularization and subsequent renormalization. Our result was obtained in the classical electromagnetic field background $A_\mu$ in the leading order of the Wigner expansion. Next, higher-order terms in the Wigner expansion would correspond to (spatial) derivative series in terms of the electromagnetic gauge field $A_\mu$ and conformal factor $\tau$. Quantum fluctuations of the electromagnetic field on top of the classical magnetic background would generate perturbative series over electromagnetic coupling $e$ at each given order of the Wigner expansion. We expect that the perturbative series would generate standard ultraviolet divergences which will be absorbed into renormalization of the gauge coupling $e$. Thus, due to renormalisability of QED, we expect that in the leading (lowest-derivative) order of the Wigner expansion the quantum corrections would lead to renormalization of the electric charge without qualitative altering of the expression for the anomalous electric current~\eq{eq:SME:2}.

We have explicitly found that for massive fermions the electric currents $J_{\measure}$ and $J_{\action}$, originating, respectively, from the anomalous symmetry breaking and from the explicit symmetry breaking, cancel each other in the leading order in the number of derivatives. Therefore, for sufficiently heavy fermions the SME should be strongly suppressed. This conclusion is in agreement with the decoupling theorem for the massive particles~\cite{Appelquist:1974tg}.

It is clear that the apparent Euclidean covariant structure of the generated current~\eq{eq:J:mu:1} is common both for the scale magnetic~\eq{eq:SME} and scale electric~\eq{eq:SEE} effects~\cite{Chernodub:2016lbo}. Therefore we believe that our conclusion may also be valid for the scale electric effect which has not been explicitly discussed in this paper.

Finally we notice that there is a potential possibility to observe the scale magnetic effect in tabletop laboratory experiments with Dirac and Weyl semimetals. These materials possess both crucial ingredients, as they host relativistic massless fermionic excitations subjected to gravitational field background. The relativistic fermions emerge naturally due to topological properties of the electronic band structure of these materials~\cite{ref:DWSM} while the emergent gravity may be induced by elastic deformations of their crystal structure~\cite{Cortijo:2015jja}. The latter effect is very common for many solid state systems~\cite{ref:applications:cond:matt}. Thus, elastically deformed topological materials may provide a useful experimental tool to study a plethora of properties~\cite{ref:book} of relativistic quantum field theory in curved space-time, including the scale magnetic effect.

\acknowledgments

M.A.Z. is grateful to LMPT, the University of Tours, where this work was initiated, for kind hospitality. The work of M.A.Z. was supported by Russian Science Foundation Grant No 16-12-10059.

\end{document}